\documentclass[aps,prd,twocolumn,showpacs,floatfix]{revtex4}
\usepackage{latexsym,hyperref,amssymb,graphicx,dcolumn,bm}
\usepackage{color} 
\newcommand{\half}{\frac{1}{2}}

\newcommand{\beq}{\begin{equation}}
\newcommand{\eeq}{\end{equation}}
\newcommand{\bea}{\begin{eqnarray}}
\newcommand{\eea}{\end{eqnarray}}
\newcommand{\bean}{\begin{eqnarray*}}
\newcommand{\eean}{\end{eqnarray*}}
\newcommand{\bei}{\begin{itemize}}
\newcommand{\eei}{\end{itemize}}
\newcommand{\ben}{\begin{enumeration}}
\newcommand{\een}{\end{enumeration}}
\newcommand{\nn}{\nonumber}
\newcommand{\ra}{\rangle}
\newcommand{\la}{\langle}
\newcommand{\ba}{\begin{array}}
\newcommand{\ea}{\end{array}}
\newcommand{\lp}{\left(}
\newcommand{\rp}{\right)}
\newcommand{\lsp}{\left[}
\newcommand{\rsp}{\right]}
\newcommand{\lcp}{\left\{}
\newcommand{\rcp}{\right\}}
\newcommand{\leqn}{\lefteqn}
\definecolor{navyblue}{rgb}{.05,0,.55}

\begin{document}
\title{Inflation in a Symmetric Vacuum}
\author{Kevin Cahill}
\email{cahill@unm.edu}
\affiliation{Department of Physics and Astronomy,
University of New Mexico, Albuquerque, NM 87131}
\date{\today}
\begin{abstract}
If in a finite universe, 
the tree-level vacuum is a symmetric
superposition of coherent states,
in each of which 
the inflaton assumes a different,
energy-minimizing 
mean value (vev),
then the resulting energy
is positive and decreases exponentially
as the volume of the 
universe increases.
This effect can drive 
inflation when that volume is small
and explain part of dark energy
when it is big, but the effect
is exceedingly tiny except at
very early times.
\end{abstract}
\maketitle
\section{The Sombrero\label{The Sombrero}}
If we quantize the fields of the standard model
in the temporal gauge \(A^0_b(x) = 0\),
then the vacuum---and all physical states---
are invariant under time-independent gauge
transformations.
So we can imagine that the
tree-level vacuum of a finite universe
is a symmetric
superposition of coherent states,
in each of which 
a scalar inflaton assumes a different
mean value (vev) that minimizes 
the inflaton potential
(and the other fields vary accordingly)\@.
The resulting energy then
is positive, 
approaching zero as the volume of the 
universe increases.
\par
In this paper, I evaluate this effect
for the simpler case of
global gauge transformations.
Integrating once over the brim
of the sombrero, I find that 
the effect can drive 
inflation when the universe is tiny
and explain a small part of dark energy
when it is huge, but that
the effect is exceedingly small except at
very early times.  
\par
I discuss coherent-state vacua
in Sec.~\ref{Coherent States}
and show that the effect vanishes
exponentially with the volume of the
universe.
I describe the symmetric coherent vacua 
and compute their energies
for theories with respectively
reflection symmetry,
U(1) symmetry, and
SU(2) symmetry in Secs.~\ref{Reflection Symmetry},
\ref{U(1) Symmetry}, and \ref{SU(2) Symmetry}\@.
\section{Coherent States\label{Coherent States}}
In what follows, we will focus on a single
mode of the inflaton---the zero-momentum mode.
In terms of this mode, 
a single real field \(\phi\) of mass \(m\) 
at time \(t=0\) is
\beq
\phi(x) = \phi = \left(\frac{1}{2mV}\right)^{1/2}
\left( a + a^\dagger \right)
\label{phi}
\eeq
and its conjugate momentum is
\beq
\pi(x) = \pi = \left(\frac{m}{2V}\right)^{1/2}
\left( - i a + i a^\dagger \right).
\label{pi}
\eeq
A coherent state~\cite{GlauberPRL1963} 
is an eigenstate of the annihilation operator \(a\)
with eigenvalue \( \alpha \)
\beq
a  | \alpha \ra = \alpha  | \alpha \ra .
\label {aalpha=alphaalpha}
\eeq
The mean values of \(\phi\) and \( \pi \)
in this state then are
\beq
\la \alpha | \phi | \alpha \ra = 
\left(\frac{1}{2mV}\right)^{1/2}
\left( \alpha + \alpha^* \right)
\label{<phi>}
\eeq
and
\beq
\la \alpha | \pi | \alpha \ra = 
\left(\frac{m}{2V}\right)^{1/2}
\left( - i  \alpha + i \alpha^*  \right).
\label{<pi>}
\eeq
\par
We will be considering 
theories in which the hamiltonian
contains both \( \pi^2 \) and 
an inflaton potential \(U\) that vanishes
for non-zero values \(\phi_0\) of the field.
In such theories,
a vacuum value
of \( \alpha \) must be real
\beq 
\alpha_0 = x_0
\label {a=x}
\eeq 
with
\beq
\phi_0 = 
\la \alpha_0 | \phi | \alpha_0 \ra = 
\la x_0 | \phi | x_0 \ra = 
\left(\frac{2}{mV}\right)^{1/2} \, x_0
\label{<xphix>}
\eeq
and \( \la x_0 | \pi | x_0 \ra = 0 \)\@.
Thus the argument \(x_0\) of a 
vacuum coherent state 
\beq
x_0 = \left(\frac{mV}{2}\right)^{1/2} \, \phi_0
\label {x0 V}
\eeq
is proportional to the square-root of the
volume of the universe.
\par
The inner product of two coherent states is
\beq
\la \beta | \alpha \ra = 
\exp\left( -\half |\alpha|^2 - 
\half |\beta|^2 + \beta^* \alpha \right),
\label {<b|a>}
\eeq
which shows that they are normalized to unity.
When both \(\alpha = x \)
and \( \beta = x' \) are real,
this inner product is
\beq
\la x' | x \ra = 
\exp\left( -\half (x - x')^2 \right).
\label {<x'|x>}
\eeq
When they also are arguments
of vacuum coherent states, their inner product
by (\ref{x0 V}) 
\beq
\la x' | x \ra = 
\exp\left( -mV(\phi_0 - \phi'_0)^2/4 \right)
\label {<f'|f>}
\eeq
vanishes exponentially fast
as the volume of the universe expands.
Thus in a spatially infinite universe,
these states and the Hilbert spaces built
upon them are orthogonal.
\section{Reflection Symmetry\label{Reflection Symmetry}}
Suppose the inflaton potential is
\beq
U = g :\left ( \phi^2 - \phi_0^2 \right)^2:
\label {Uref}
\eeq 
in which the colons denote normal ordering,
that is, \(a\)'s appear to the right of
all \(a^\dagger\)'s.
There are then two vacuum coherent states 
\( | \pm x_0 \ra \)
with \(x_0\) given by (\ref{x0 V})
for which the mean value of the potential vanishes
\beq
\la x_0 | U | x_0 \ra = \la -x_0 | U | -\!x_0 \ra = 0.
\label {<+-U+->=0}
\eeq
\par
Under the reflection \( \phi \to - \phi \),
the states \( | \pm x_0 \ra \) are interchanged.
The symmetric state
\beq
| Sx_0 \ra = 
N \left( | x_0 \ra
+ | -x_0 \ra \right)
\label {x+-x}
\eeq
is invariant under a reflection
and is normalized when
\beq
N = \left[ 2 \left( 1 
+ e^{-2x_0^2} \right) \right]^{-\half}.
\label {N}
\eeq
\par
Because the matrix elements
\(\la \pm x_0 | : \phi^n : | \mp x_0 \ra\)
vanish when \(n\) is a non-negative integer,
the  mean value of the inflaton potential 
in the symmetric state \( | Sx_0 \ra \) is
\beq
\la Sx_0 | U | Sx_0 \ra = 
\frac{g \, \phi_0^4 }{(1 + e^{mV\phi_0^2})}.
\label {<U>}
\eeq
\par
This energy is huge when \(mV\phi_0^2\) is tiny
and tiny when \(mV\phi_0^2\) is huge.
So although unrealistic,
it has the qualitative form 
to drive inflation in a baby universe
and to accelerate the expansion
in a mature universe.
Unfortunately, the epoch
during which \(mV\phi_0^2\) is small
is very brief, and the contribution
to dark energy when \(mV\phi_0^2\) is big
is exponentially suppressed.
\section{U(1) Symmetry\label{U(1) Symmetry}}
We now consider the more realistic case
of a complex inflaton 
\(\phi = ( \phi_1 + i \phi_2)/\sqrt{2}\)
made as usual of two real fields of equal mass.
We'll take the inflaton potential to be
\beq
U_1 = g :\left ( \phi^\dagger \phi - \phi_0^2 \right)^2:
\label {U1ref}
\eeq 
in which \(\phi_0 > 0\)\@.
Now the vacuum coherent states are
direct-product states
\( | x_1, x_2 \ra \) with
\beq
x_1^2 + x_2^2 = \left( \frac{mV}{2} \right)
\left( \phi_{10}^2 + \phi_{20}^2  \right)
= m V \phi_0^2.
\label {x1^2+x_2^2} 
\eeq
Their inner products by (\ref{<x'|x>}) are
\beq
\la x'_1, x'_2 |  x_1, x_2 \ra = 
e^{- (x_1 - x_1')^2/2} e^{ - (x_2 - x'_2)^2/2 }.
\label {<x1x2|x1'x2'>}
\eeq
We may denote these states by an angle \(\theta\)
\beq
| x_1, x_2 \ra = | \theta \ra
\label {th}
\eeq
with
\beq
x_1 + i x_2 = \sqrt{mV} \, \phi_0 e^{i \theta}.
\label {eith}
\eeq
In terms of angular variables,
the inner product (\ref{<x1x2|x1'x2'>}) then is
\beq
\la \theta' | \theta \ra = 
\exp\left\{-2mV\phi_0^2 \sin^2[(\theta - \theta')/2]
\right\}.
\label {<th'|th>}
\eeq
\par
The inflaton vacuum \( |S\phi_0\ra \) 
\beq
|S\phi_0\ra = N_1 \int_0^{2\pi} \!d\theta \, | \theta \ra
\label {Sphi0}
\eeq
is symmetric; it is normalized when
\beq
N_1 = 1/\sqrt{8\pi W(y_0)}
\label {N1norm}
\eeq
where \(W(y)\) is
\beq
W(y) = \int_0^{\pi/2} \!\! dx \,
\exp\left(-y\sin^2x\right)
\label {W(y)}
\eeq
and \(y_0 = 2mV \phi_0^2\)\@.
The integral formula
\beq
\int_0^{\pi/2} \!\! dx \, \sin^{2n}x =
\frac{(2n-1)!!}{(2n)!!} \, \frac{\pi}{2}
\label {intForm}
\eeq
lets us write \( W(y) \) as the power series
\beq
W(y) = \frac{\pi}{2} \sum_{n=0}^\infty 
\frac{(-y)^n}{n!} \, 
\frac{(2n-1)!!}{(2n)!!}.
\label {W(y)sum}
\eeq
\par
The complex field \(\phi\) is
\beq
\phi = \left(\frac{1}{2\sqrt{mV}}\right)
\left[a_1 + a_1^\dagger + i ( a_2 + a_2^\dagger) \right]
\label {phic}
\eeq
and so some of its coherent-state 
matrix elements are
\bea
\la \theta' | : (\phi^\dagger)^n \phi^m : | \theta \ra
& = & \la \theta' | \theta \ra \phi_0^{n+m} 
\left( \frac{e^{-i\theta'} + e^{-i\theta}}{2} \right)^n 
\nn\\
& & \times 
\left( \frac{e^{i\theta'} + e^{i\theta}}{2} \right)^m.
\label {<th'|f*nfmth>}
\eea
The matrix element of the inflaton potential thus is
\beq
\la \theta' | U_1 | \theta \ra = g \la \theta' | \theta \ra 
\phi_0^4 \sin^4 [(\theta - \theta')/2].
\label {<thi|U|th>}
\eeq
The mean value of this potential 
in the symmetric state \( |S\phi_0\ra \)
then is
\beq
\la S\phi_0 |U_1|S\phi_0\ra = g \phi_0^4 
W(y_0)^{-1} W''(y0).
\label {d2W}
\eeq
\par
For \(y \ll 1 \),
we may use the power series (\ref{W(y)sum})
to write
\beq
W(y) \approx \frac{\pi}{2} \left( 1 - \frac{y}{2}
+ \frac{3y^2}{16} - \frac{5y^3}{96} 
+ \frac{35y^4}{3072} + \dots \right)
\label {Wy<1}
\eeq
and
\beq
W''(y) \approx \frac{\pi}{16} 
\left( 3 - \frac{5}{2} y 
+ \frac{ 35 y^2}{32} + \dots \right).
\label {W''y<1}
\eeq 
So to lowest order in \(y_0 = 2mV \phi_0^2 \ll 1 \),
the mean value of the inflaton potential 
in the symmetric state \( |S\phi_0\ra \) is
\beq
\la S\phi_0 |U_1|S\phi_0\ra \approx
\frac{3g}{8} \phi_0^4
\left( 1 - \frac{2}{3} m V \phi_0^2 \right),
\label {U1y<1}
\eeq
which might drive inflation
in an early, tiny universe
while \( mV\phi_0^2 \) is small.
But that epoch is very brief.
\par
In the other limit, \( y \to \infty\),
we may use the approximations
\beq
W(y) \approx \int_0^\infty e^{-yx^2} dx 
= \half \sqrt{\frac{\pi}{y}}
\label {Wyinf}
\eeq
and
\beq
W''(y) \approx \int_0^\infty x^4 e^{-yx^2} dx 
= \frac{3}{8y^2} \sqrt{\frac{\pi}{y}}
\label {W''yinf}
\eeq
to estimate the dark-energy density as
\beq
\la S\phi_0 |U_1|S\phi_0\ra \approx
\frac{3g}{16 m^2 V^2}
\label {U1DE}
\eeq
as \(y_0 = 2mV \phi_0^2 \to \infty\)\@.
This energy density would be a negligibly small
contribution to dark energy.
\section{SU(2) Symmetry\label{SU(2) Symmetry}}
The inflaton of the standard model
has two complex components
\beq
h = \frac{1}{\sqrt{2}} \lp 
\ba{c} \phi_1 + i \phi_2 \\
\phi_3 + i \phi_4 \ea
\rp
\label {h}
\eeq
in which \(\phi_i\) 
are real scalar fields of mass \(m\)\@.
The inflaton potential is then
\beq
U_2 = g : \lp h^\dagger h - \phi_0^2 \rp^2 :.
\label {U2}
\eeq
The vacuum coherent states now
are direct-product states
\( |x_1,x_2,x_3,x_4\ra \) whose
arguments lie on the surface
of a three-dimensional sphere
of radius \(\sqrt{mV} \phi_0 \)
in four-space:
\bea
x_1^2+x_2^2+x_3^2+x_4^2 & = &
\lp \frac{mV}{2} \rp
\lp \phi_{10}^2+\phi_{20}^2+\phi_{30}^2+\phi_{40}^2 \rp 
\nn\\ & = & mV \lp |h_1|^2 + |h_2|^2 \rp
= mV \phi_0^2.
\label {xifi}
\eea
By an extension of Eq.~(\ref{<x1x2|x1'x2'>}),
their inner product is
\beq
\la x'_1,x'_2,x'_3,x'_4 | x_1,x_2,x_3,x_4\ra
= e^{-(x - x')^2/2}
\label {<x'1x'4|x1x4>}
\eeq
in which
\beq
(x - x')^2 = \sum_{i=1}^4 (x_i - x'_i)^2.
\label {(x-x')^2}
\eeq
In Hopf's coordinates,
\bea
x_1 + i x_2 & = & 
\lp \sqrt{mV} \phi_0 \rp \, e^{i\theta_1} \sin \eta 
\nn\\
x_3 + i x_4 & = & 
\lp \sqrt{mV} \phi_0 \rp \, e^{i\theta_2} \cos \eta 
\label {Hopf}
\eea
the symmetric state \( | S_2 \phi_0 \ra \) is
\beq
| S_2 \phi_0 \ra = N_2 \int_0^{2\pi} \!\! d\theta_1
\int_0^{2\pi} \!\! d\theta_2 
\int_0^{\pi/2} \!\! d \eta \, \sin 2\eta \,
| \theta_1, \theta_2, \eta \ra
\label {S2f0}
\eeq
and the inner product (\ref{<x'1x'4|x1x4>}) is
\bea
\la \theta'_1, \theta'_2, \eta' 
| \theta_1, \theta_2, \eta \ra
& = & \exp \lcp -mV\phi_0^2 \lsp 1  \right. \right. 
\\ & & 
- \cos(\theta_1 - \theta'_1) \sin \eta \sin \eta'
\nn\\ & & \left. \left.
- \cos(\theta_2 - \theta'_2) \cos \eta \cos \eta'
\rsp \rcp . \nn
\label {th'th'e|th th e}
\eea
The normalization constant \(N_2\) is
\beq
N_2 = 1/\lp 2 \pi \sqrt{W_2(y_0)} \rp
\label {N2=}
\eeq
where \( W_2(y) \) is the integral
\bea
W_2(y) & = & e^{-y} \! \int_0^{2\pi} \!\!\! d\theta_1
\int_0^{2\pi} \!\!\! d\theta_2 
\int_0^{\pi/2} \!\!\!\!\! d \eta 
\int_0^{\pi/2} \!\!\!\!\! d \eta'
\sin 2\eta \sin 2\eta' \nn\\ 
& & \times 
e^{y \cos \theta_1 \sin \eta \sin \eta'} 
e^{y \cos \theta_2 \cos \eta \cos \eta'}
\label {W2=}
\eea
and \( y_0 = mV \phi_0^2 \)\@.
\par
Some of the coherent-state matrix elements
of normally ordered monomials
of the inflaton components are
\bea
\leqn{
\la \theta'_1, \theta'_2, \eta' |
:\lp h^\dagger_1 \rp^n h_1^j 
\lp h^\dagger_2 \rp^k h_2^\ell :
| \theta_1, \theta_2, \eta \ra = \phi_0^{n+j+k+\ell}} \nn\\
& & \times \la \theta'_1, \theta'_2, \eta' 
| \theta_1, \theta_2, \eta \ra
\lp \frac{e^{-i\theta_1} \sin \eta + 
e^{-i\theta'_1} \sin \eta' }{2} \rp^n \times
\nn\\
& & 
\lp \!\frac{e^{i\theta_1} \sin \eta + 
e^{i\theta'_1} \sin \eta' }{2} \rp^j
\!\!\lp \!\frac{e^{-i\theta_2} \cos \eta + 
e^{-i\theta'_2} \cos \eta' }{2}\! \rp^k
\nn\\
& & \times \lp \frac{e^{i\theta_2} \cos \eta + 
e^{i\theta'_2} \cos \eta' }{2}\! \rp^\ell.
\eea
So the corresponding matrix element 
of the inflaton potential (\ref{U2}) 
is
\bea
\leqn{
\la \theta'_1, \theta'_2, \eta' | U_2
| \theta_1, \theta_2, \eta \ra = \frac{\phi_0^4}{4}
\la \theta'_1, \theta'_2, \eta' |
\theta_1, \theta_2, \eta \ra \times} \\
& & \!\! \lsp 1
- \cos(\theta_1 - \theta'_1) \sin \eta \sin \eta'
- \cos(\theta_2 - \theta'_2) \cos \eta \cos \eta'
\rsp^2 \nn
\eea
So the mean value of the inflaton potential
(\ref{U2}) in the symmetric vacuum (\ref{S2f0}) is
\beq
\la S_2 \phi_0 | U_2 | S_2 \phi_0 \ra = 
\frac{g\phi_0^4}{4} \, \frac{W''_2(y_0)}{W_2(y_0)}.
\label {d2W2}
\eeq
\par
We may expand the quartic integral (\ref{W2=})
in powers of \(y\), do the
integrals over \(\theta_1\), \(\theta_2\),
\(\eta\), and \(\eta'\), and write
\(W_2(y)\) as the double sum 
\beq
W_2(y) = 8 \pi^2 e^{-y} \!\! \sum_{j,k=0}^\infty
\frac{j!(2j-1)!!}{(2j)!(2j)!!}
\frac{k!(2k-1)!!}{(2k)!(2k)!!} 
\frac{y^{2j+2k}}{(j+k+1)!}
\eeq
which is useful in the \(y \to 0\) limit
where
\beq 
W_2(y) \approx  8 \pi^2 e^{-y}
\lp 1 + \frac{y^2}{4} + \frac{y^4}{96} \rp.
\label {W2approxTinyV}
\eeq
So in an early, tiny universe the vacuum energy is
\beq
\la S_2 \phi_0 | U_2 | S_2 \phi_0 \ra \approx 
\frac{3g\phi_0^4}{8} \, 
\lp 1 - \frac{2mV\phi_0^2}{3} 
+ \frac{(mV\phi_0)^3}{9} \rp.
\label {W2tinyV}
\eeq
Unfortunately,
the period during which \( mV\phi_0^2 < 1 \)
is very brief---it is when the radius
of the universe was small compared
to \(1/m\) and to \(1/\phi_0\)\@.
\par
In the \(y \to \infty\) limit,
\beq
W_2(y) \approx \frac{16\pi\sqrt{2\pi}}{y^{5/2}}
\label {W2bigV}
\eeq
and so the dark-energy density is
\beq
\la S_2 \phi_0 | U_2 | S_2 \phi_0 \ra \approx 
\frac{35}{16} \, \frac{g}{m^2 V^2},
\label {DE2bigV}
\eeq
which would be a negligibly small
contribution to dark energy.
\section{Conclusion\label{Conclusion}}
A globally gauge-invariant coherent-state vacuum might
explain inflation in the very early
universe, but I can't imagine a model 
in which this effect could cause 40 e-foldings;
it could contribute to dark energy
in a big universe, but only negligibly.
\par
In this paper, I have considered for simplicity
only global gauge transformations
and the zero-momentum mode of the inflaton.
By path-integrating over
all time-independent gauge transformations 
in a model with a more realistic 
inflaton~\cite{Allahverdi2007},
one might arrive at an effect
that numerically is more impressive. 
\bibliography{physics,math}
\end{document}